\documentclass{article}
\usepackage[utf8]{inputenc}
\usepackage{enumitem}
\usepackage{listings}
\usepackage[T1]{fontenc}
\usepackage{multirow}
\usepackage[utf8]{inputenc}

\usepackage{spconf,amsmath,graphicx}

\usepackage{microtype}
\usepackage{graphicx}

\usepackage{subcaption}

\usepackage{booktabs} 
\usepackage{xcolor} 
\usepackage{pgfplots,pgfplotstable}

\usepackage{hyperref}
\usepackage{url}

\usepackage{amsmath}
\usepackage{amssymb}
\usepackage{mathtools}
\usepackage{amsthm}
\usepackage{bm}
\usepackage[capitalize,noabbrev]{cleveref}

\definecolor{forestgreen}{rgb}{0.13, 0.55, 0.13}
\definecolor{fulvous}{rgb}{0.86, 0.52, 0.0}
\definecolor{glaucous}{rgb}{0.38, 0.51, 0.71}
\definecolor{lava}{rgb}{0.81, 0.06, 0.13}
\definecolor{buff}{rgb}{0.94, 0.86, 0.51}
\definecolor{chromeyellow}{rgb}{1.0, 0.65, 0.0}
\definecolor{brightube}{rgb}{0.82, 0.62, 0.91}

\hypersetup{
    pdfborderstyle={/S/U/W 1}, 
    linkbordercolor=red,       
    citebordercolor=green,     
    filebordercolor=magenta,   
    urlbordercolor=cyan        
}

\pgfplotsset{scaled y ticks=false, scaled x ticks=false}

\theoremstyle{plain}

\theoremstyle{definition}

\theoremstyle{remark}

\usepackage{xspace}

\usepackage[textsize=tiny]{todonotes}

\usepackage{siunitx}

\usepackage{float}
\usepackage{adjustbox}

\title{Scaling NVIDIA's Multi-speaker Multi-lingual TTS Systems with Zero-Shot TTS to Indic Languages}

\name{
    \begin{tabular}{c c c c c}
    Akshit Arora & Rohan Badlani & Sungwon Kim & Rafael Valle & Bryan Catanzaro
    \end{tabular}
}

\address{NVIDIA}
\begin{document}

\maketitle

\section{Abstract}
\vspace*{-0.75\baselineskip}
In this paper, we describe the TTS models developed by NVIDIA for the MMITS-VC (Multi-speaker, Multi-lingual Indic TTS with Voice Cloning) 2024 Challenge.  In Tracks 1 and 2, we utilize RAD-MMM~\cite{badlani2023multilingual} to perform few-shot TTS by training additionally on 5 minutes of target speaker data. In Track 3, we utilize P-Flow~\cite{kim2023pflow} to perform zero-shot TTS by training on the \emph{challenge dataset} as well as external datasets. We use HiFi-GAN~\cite{kong2020hifi} vocoders for all submissions. RAD-MMM performs competitively on Tracks 1 and 2, while P-Flow ranks first on Track 3, with mean opinion score (MOS) 4.4 and speaker similarity score (SMOS) of 3.62.

\vspace*{-1.0\baselineskip}
\section{Introduction}
\vspace*{-0.75\baselineskip}
The MMITS-VC 2024 Challenge\footnote{\url{https://sites.google.com/view/limmits24/home}} is organized as a part of ICASSP's Signal Processing Grand Challenge in 2024. It aims at the development of multi-speaker multi-lingual TTS (TTS) systems capable of performing zero-shot TTS. A \emph{challenge dataset} of 7 languages (Hindi, Telugu, Marathi, Bengali, Chhattisgarhi, English and Kannada), with 2 speakers per language, is made available to the participants. For Tracks 1 and 2, additional 5 minutes of data from the target evaluation speakers is provided to promote few-shot TTS. Participants build TTS models and share the generated audio samples for mono and cross lingual evaluation of samples through extensive listening tests. 

There has been incredible progress in the quality of Text-To-Speech (TTS) models, specially in zero-shot TTS with large-scale neural codec autoregressive language models. Unfortunately, these models inherit several drawbacks from earlier autoregressive models: they require collecting thousands of hours of data, rely on pre-trained neural codec representations, lack robustness, and have very slow sampling speed. These issues are not present in the models we propose here. 


Our goal, in Tracks 1 and 2, is to create a multi-lingual TTS system that can synthesize speech in any target language (with a target language's native accent) for any speaker seen by the model. We use RAD-MMM~\cite{badlani2023multilingual} to disentangle attributes such as speaker, accent and language, such that the model can synthesize speech for the desired speaker, and the desired language and accent, without relying on any bi-lingual data.

In Track 3, our goal is to create a multi-lingual TTS system that synthesizes speech in any seen target language given a speech prompt. We use P-Flow~\cite{kim2023pflow}, a fast and data-efficient zero-shot TTS model that uses speech prompts for speaker adaptation. 

\vspace*{-1.0\baselineskip}
\section{Method}
\vspace*{-0.75\baselineskip}
\label{Dataset and Preprocessing}
\subsection{Dataset and Preprocessing}
\vspace*{-0.7\baselineskip}
We reprocess the provided speech data (challenge and few-shot datasets) with the approach described in our previous submission to the LIMMITS 2023 challenge~\cite{vani}. During pre-processing, we remove empty audio files and clips with duplicate transcripts, trim leading and trailing silences, and normalize audio volume. This results in the  \emph{\emph{challenge dataset}} that we use in all the tracks. Statistics for the  \emph{\emph{challenge dataset}} are captured here \url{https://bit.ly/mmits24_nvidia}.

In addition to the \emph{challenge dataset}, we use LibriTTS~\cite{zen2019libritts} and VCTK~\cite{Veaux2017CSTRVC} for tracks where external datasets are allowed by the challenge rules.

\vspace*{-0.75\baselineskip}
\subsection{Tracks 1 and 2: Few-shot TTS with RAD-MMM}
\vspace*{-0.7\baselineskip}
Our goal is to develop a model for multilingual synthesis in the languages of interest with the ability of cross-lingual synthesis for a (seen) speaker of interest. Our dataset comprises of each speaker speaking \emph{only one language} and hence there are correlations between text, language, accent and speaker within the dataset. Recent work on RAD-MMM~\cite{badlani2023multilingual} tackles this problem by proposing several disentanglement approaches. Following RAD-MMM, we use deterministic attribute predictors to separately predict fine-grained features like fundamental frequency (F0) and energy given text, accent and speaker.

In our setup, we leverage the text pre-processing, shared alphabet set and the accent-conditioned alignment learning mechanism proposed in RAD-MMM. We consider language to be \emph{implicit in the phoneme sequence}, whereas the information captured by the accent should explain the fine-grained differences between \emph{how phonemes are pronounced in different languages}. 

\vspace*{-1.0\baselineskip}
\subsubsection{Track $1$} 
\vspace*{-0.6\baselineskip}
We train RAD-MMM on the \emph{\emph{challenge dataset}} described in Section~\ref{Dataset and Preprocessing}. Since the few-shot dataset is very small (5 mins per speaker), to avoid overfitting on small-data speakers, we fine-tune the trained RAD-MMM model with both challenge and few-show data for 5000 iterations and batch size 8. 

\vspace*{-1.0\baselineskip}
\subsubsection{Track $2$} 
\vspace*{-0.6\baselineskip}
We train RAD-MMM on the \emph{challenge dataset}, LibriTTS and VCTK (excluding target evaluation speakers from the few-shot dataset, following challenge guidelines). Even though the additional datasets (LibriTTS and VCTK) are English only, they contain many speakers, thus helping the model generalize better. Similarly to Track 1, we avoid overfitting by fine-tuning this RAD-MMM model on challenge and few-shot data for 5000 iterations and batch size 8. 

\vspace*{-0.5\baselineskip}
\subsection{Track 3: Zero-shot TTS with P-Flow}
\vspace*{-0.7\baselineskip}
In Track 3, our goal is to perform zero-shot TTS for multiple languages using reference data for speakers unseen during training. The \emph{challenge dataset} contains speech samples from two speakers for each language, with at most 14 speakers. To achieve zero-shot TTS for new speakers, it is necessary to learn to adapt to a variety of speakers. Therefore, we additionally utilize the English multi-speaker dataset, LibriTTS, to alleviate the shortage of speakers in the Indic languages of the \emph{challenge dataset}.

Recently, P-Flow~\cite{kim2023pflow} has demonstrated strong ability for zero-shot TTS in English by introducing speech prompting for speaker adaptation. P-Flow performs zero-shot TTS using only 3 seconds of reference data for the target speaker. We expand this capability for cross-lingual zero-shot, extending the model's zero-shot TTS ability from its original language, English, to include seven additional Indic languages. To achieve this, we modify P-Flow's decoder architecture, extend the training data to include the \emph{challenge dataset}, and use RAD-MMM's~\cite{badlani2023multilingual} text pre-processing to expand to other languages.

We train this modified version of P-Flow on both LibriTTS and the \emph{challenge dataset}s, filtering out audio samples shorter than $3$ seconds. In total, we utilize $964$ hours of speech data from $2287$ speakers, comprising $14$ speakers from the \emph{challenge dataset} and an additional $2273$ speakers from LibriTTS. We train P-Flow on $8$ A$100$ GPUs, each with a batch size of $8$ samples, achieving the same effective batch size of $64$ as in the original P-Flow paper. We follow the same training details as the P-Flow paper.

During inference, we use Euler's method for sampling with the default setup of P-Flow, with a classifier-free guidance scale of 1 and 10 sampling steps. For evaluation, we only use $3$ seconds of speech from the given target speaker for zero-shot TTS. Specifically, we randomly select one sample corresponding to 3-4 seconds from each target speaker's reference sample and only crop the first 3 seconds for use. Note that we do not use the transcript for the reference sample, so the model does not receive information about the language being used.

\vspace*{-0.5\baselineskip}
\subsection{Vocoder}
\vspace*{-0.7\baselineskip}
For Track 1, we use the \emph{challenge dataset} to train a mel-conditioned HiFi-GAN vocoder~\cite{kong2020hifi} model for both mono and cross lingual scenarios. For Track 3, we use the challenge and VCTK~\cite{Veaux2017CSTRVC} datasets (excluding target evaluation speakers from few-shot dataset, following challenge guidelines) to train a HiFi-GAN vocoder~\cite{kong2020hifi} model from scratch. For Track 2, we further finetune the vocoder trained for Track 3 on the few-shot training dataset provided by challenge organizers.

\vspace*{-1.0\baselineskip}
\section{Results}
\vspace*{-0.75\baselineskip}
The official results and competition leaderboard is available at \url{https://sites.google.com/view/limmits24/results}. 



\vspace*{-1.0\baselineskip}
\section{Conclusion}
\vspace*{-0.75\baselineskip}
This paper presents the TTS systems developed by NVIDIA for the MMITS-VC Challenge 2024. We use RAD-MMM for few-shot TTS scenarios (Tracks 1 and 2) because of its ability to disentangle speaker, accent and text for high-quality multilingual speech synthesis without relying on bi-lingual data. We use P-Flow for zero-shot TTS scenario (Track 3) as alongside HiFi-GAN based vocoder models due to its ability to perform zero-shot TTS with a short 3 seconds audio sample. The challenge evaluation shows that RAD-MMM performs competitively on Tracks 1 and 2, while P-Flow ranks first on Track 3. 



\vspace*{-1.0\baselineskip}
\bibliographystyle{IEEEbib}
\bibliography{main}\label{sec:references}

\end{document}